\documentstyle[11pt,aaspp4,flushrt]{article}

\begin{document}

\title{The Globular Cluster M54 and the \\Star Formation History of 
the Sagittarius Dwarf Galaxy}
\author{Andrew C. Layden\altaffilmark{1,2}}
\affil{Dept. of Astronomy, University of Michigan, Ann Arbor, MI 48109-1090}
\authoremail{layden@astro.lsa.umich.edu}
\author{Ata Sarajedini\altaffilmark{1}}
\affil{Kitt Peak National Observatory, P.O. Box 26732, Tucson, AZ 85726-6732}
\authoremail{ata@noao.edu}

\bigskip
\bigskip
\begin{center}
Accpeted for publication in Astrophysical Journal Letters.
\end{center}

\altaffiltext{1}{Hubble Fellow.}

\altaffiltext{2}{Observations obtained while a staff member of the Cerro
Tololo Inter-American Observatory, one of the National Optical
Astronomy Observatories (NOAO).  NOAO is operated by the Association
of Universities for Research in Astronomy, Inc., under cooperative
agreement with the National Science Foundation.}

\begin{abstract}

We present a deep color-magnitude diagram in the $VI$ passbands of the
globular cluster M54, a member of the Sagittarius dwarf galaxy.  The
data extend below the cluster's main sequence turn-off, allowing us to
estimate the cluster's age.  We find that M54 is 0.5--1.5 gigayears
older than the Galactic globulars M68 and M5.  In absolute terms, the
age is comparable to the published age estimates of the other member
clusters Arp 2 and Terzan 8, but is significantly older than the
member cluster Terzan 7.  An age estimate of the Sagittarius field
population relative to M54 suggests that M54 is $\gtrsim$3 Gyr older
than the field.  We discuss briefly the star formation history of the
Sagittarius dwarf galaxy.

\end{abstract}

\keywords{galaxies: individual (Sagittarius) --- galaxies: star
clusters --- galaxies: stellar content --- globular clusters:
individual (NGC 6715) --- stars: Population II}

\section{Introduction}

\markcite{igi94}Ibata, Gilmore, \& Irwin (1994, hereafter referred to
as IGI) announced the discovery of a gas-poor galaxy lying behind the
bulge of the Milky Way at a distance of $\sim$24 kpc.  They estimated
the overall stellar content of this galaxy (hereafter referred to as
the Sagittarius dwarf galaxy, or Sgr) to be similar in age,
metallicity, and total luminosity to the Fornax dwarf spheroidal
galaxy.  They also noted that four previously-known globular clusters
(M54, Arp 2, Ter 7, and Ter 8) had roughly the correct distances,
positions on the sky, and radial velocities to be members of this
galaxy.

The presence of a galaxy so nearby provides an unparalleled
opportunity for detailed study of stellar populations, chemical
enrichment history, and galaxy formation and evolution.
\markcite{igi97}Ibata {\em et~al.} (1997) provide an excellent review
of the ensuing research on Sgr.  The results of several of these
studies have increased the likelihood that the four globular clusters
are members of Sgr.  In particular, \markcite{dca95}Da~Costa \&
Armandroff (1995) presented new abundances and radial velocities for
the four globular cluster candidates; their velocities agree well with
the systemic velocity of Sgr (\markcite{igi94}IGI).  The new
isodensity map of \markcite{igi97}Ibata {\em et al.} (1997) confirms
that Sgr extends over the three outlying clusters, Arp 2, Ter 7, and
Ter 8.

Table 1 summarizes our current knowledge concerning the ages of Sgr
and its globular clusters.  The difficulties in comparing absolute
ages of stellar systems are well-known (e.g., \markcite{cha95}Chaboyer
1995), so we have used ages, when available, from the self-consistent
work of \markcite{cds96}Chaboyer {\em et~al.}  (1996, hereafter
referred to as CDS).  We use the ages from column 2 of their Table 3,
which employ an $M_V(RR)-[Fe/H]$ relation nearly identical to that of
\markcite{ldz90}Lee {\em et~al.}  (1990, hereafter referred to as
LDZ).  \markcite{cds96}CDS do not include ages for the Sgr field
population, so the estimates for Sgr listed in Table 1 are less
homogeneous.  The confusion is compounded by the uncertainty in the
metallicity of Sgr; the dependence of derived age on assumed
metallicity is well known (e.g., \markcite{cds96}CDS).  Still, the
studies agree on an age between 10 and 14 Gyr for the dominant
population of Sgr.

The only Sgr globular cluster candidate currently without accurate
main sequence turn-off (MSTO) photometry and an age estimate is M54
(NGC 6715).  \markcite{sl95}Sarajedini \& Layden (1995, hereafter
referred to as SL95) presented a CCD color-magnitude diagram (CMD) of
the bright stars in M54 ($M_V \lesssim +2.0$).  They found M54 to be
metal poor ($[Fe/H] = -1.79$ dex) and to have a blue horizontal branch
typical of old Galactic globular clusters.  The recent photometry of
M54 by \markcite{mbc97}Marconi {\em et~al.} (1997) goes deeper, but
the photometric errors and incompleteness at the magnitude of the MSTO
preclude an accurate age analysis.

M54 is by far the most luminous of the four Sgr clusters
(\markcite{sl95}SL95), and perhaps the most secure candidate for
membership in Sgr, since it lies in the highest density region of that
galaxy (\markcite{igi94}IGI) and has a distance (\markcite{sl95}SL95)
and radial velocity (\markcite{dca95}Da~Costa \& Armandroff 1995)
which correspond very well with those of Sgr.  \markcite{sl95}SL95
have speculated that M54 is the ``nucleus'' of Sgr, akin to the nuclei
found in many dwarf elliptical galaxies.  Clearly, we cannot have a
complete understanding of the star formation history of Sgr without
information about the age of M54.

\section{Observations, Reductions, and the CMD}

In order to obtain photometry to the MSTO, we secured deep images
of M54 using the CTIO 4.0-m telescope at prime focus during the nights
of 1995 June 28 and 29.  We employed the Tek\#4 2048$^2$ pixel
CCD, which provided a scale of 0.43 arcsec pixel$^{-1}$.  The median
seeing was 1.2 arcsec.  

We measured instrumental stellar magnitudes on each frame using the
DoPHOT photometry program (\markcite{sms93}Schechter, Mateo, \& Saha
1993), and transformed them directly to the $VI$ magnitude system of
\markcite{sl95}SL95 using the large number of stars common to both
data sets.  Photometry from the five $VI$ frame pairs with the best
seeing were assembled, and mean magnitudes and errors (standard errors
of the mean) were computed for each star detected in three or more
frame pairs.  The details of the reduction procedure and the
photometric data for the resulting 26,485 stars are presented in
\markcite{ls97}Layden \& Sarajedini (1997).  Comparisons show that
these data are accurately tied to the \markcite{sl95}SL95 photometric
system.

 
Figure 1 presents the $VI$ CMDs for (a) all the M54 stars with high
quality data, and (b) all the high quality M54 stars located between
2.5 and 4.3 arcmin from the cluster center.  In these panels, the
curves are the fiducial red giant branches (RGBs) of M54 and Sgr
derived by \markcite{sl95}SL95.  These curves, together with the CMDs
of Sgr and a foreground bulge control field by \markcite{muskk}Mateo
{\em et~al.}  (1995, hereafter referred to as MUSKKK), facilitate the
interpretation of our CMD.

The curve on the left is the M54 RGB fiducial; it guides the eye
faintward to where the M54 RGB becomes well populated.  The M54 RGB
turns blueward onto the subgiant branch (SGB) at ($V-I$, $V$) = (0.9,
20.9) mag, and merges with a column of points $\sim$0.2 mag blueward
of this.  As we will see, this column represents the superimposed MSTO
regions of M54 and Sgr.

The curve on the right is the Sgr RGB fiducial.  The lower RGB of Sgr
is not as well populated as that of M54 in this figure, but there
appears to be an excess of points roughly parallel to the M54 RGB
which terminates at ($V-I$, $V$) = (1.0, 20.9) mag, and which
presumably turns blueward onto the SGB at this point (see
\markcite{muskk}MUSKKK).  The MSTO of Sgr in the \markcite{muskk}MUSKKK
field occurs at ($V-I$, $V$) $\approx$ (0.75, 21.4) mag, the same
region as the MSTO of M54 in our data.  The plume of stars at $V-I$ =
0.75 and $20.2 < V < 20.9$ mag corresponds to the young (4 Gyr) Sgr
population discovered by \markcite{muskk}MUSKKK.

Other prominent sequences in Figure 1a include the M54 blue HB
(\markcite{sl95}SL95), the Sgr red HB clump (\markcite{muskk}MUSKKK,
\markcite{sl95}SL95), and a population of blue stragglers or very
young stars belonging either to M54 or Sgr ($0.2 <V-I < 0.6$, $19 < V
< 21$ mag).  The \markcite{muskk}MUSKKK bulge control field coincides
well with the column of stars at $V-I \approx 0.9$ and $V < 19.5$ mag
which sweeps redward at fainter magnitudes across the M54 and Sgr
lower RGBs.  Thus most of the scatter with $V > 20$ and $V-I > 1.0$
mag is attributed to the foreground bulge.

One important qualitative statement about the relative ages of M54 and
Sgr can be made at this point.  The MSTOs of these populations appear
to be coincident at $V-I \approx 0.75$ mag.  The reddenings are
identical since the populations lie in the same field.  Yet the
metallicity of M54 is {\em at least} 0.5 dex lower than that of Sgr,
so for the MSTOs to coincide, M54 must be {\em older} than Sgr.

\section{Comparison with Cluster Fiducial Sequences}

A simple estimate of the age of M54 can be made by directly comparing
our photometry with the fiducial sequences of other clusters.  High
quality $VI$ CCD photometry exists in the literature for M68 ($[Fe/H]
= -2.09$) and M5 ($[Fe/H] = -1.40$), which bracket M54 in metallicity.
Figure 2 shows our data plotted with the fiducial sequence of M68 from
\markcite{wal93}Walker (1994, dashed line) and the fiducial of M5 from
\markcite{sbs96}Sandquist {\em et~al.} (1996, solid line).  All the
cluster data were registered to the observational HR Diagram using the
$V(HB)$, $[Fe/H]$, and $E(V-I)$ values given in Table 2 along with the
relation $M_V(RR) = 0.17[Fe/H] + 0.82$ (\markcite{ldz90}LDZ).

The fiducial sequence comparison reveals that the age of M54 is
comparable to those of M68 and M5.  Since M54 is almost exactly
between M68 and M5 in metallicity, one expects the data for M54 to lie
midway between the M68 and M5 fiducials.  However, the M54 data,
particularly for the SGB, appears to be skewed slightly toward the M5
fiducial.  This suggests that M54 may be slightly older than M68 or
M5.  In the next section, we will quantify this age difference.
 
\section{Isochrone Fitting}
 
Another method for measuring globular cluster ages is isochrone
fitting.  Figure 3 shows the Revised Yale Isochrones (RYI; Green,
Demarque, \& King 1987) for $Y=0.23$, $[Fe/H] = -1.50$, and ages of
10--18 Gyr, superimposed on the data from Figure 1b.  The isochrones
were shifted to the observed plane using the $V(HB)$ and $E(V-I)$
values listed in Table 2, and the \markcite{ldz90}LDZ relation between
$M_V(RR)$ and $[Fe/H]$ (assumed for consistency with the RYI, see
\markcite{kea88}King {\em et al.} 1988).  The ridge-line of M54 SGB
stars suggests an age of 13--14 Gyr.  The 12 Gyr isochrone forms an
envelope about the MSTO points, setting a hard lower limit for the age
of M54 under the stated assumptions.

We used isochrones with $[Fe/H] = -1.50$ because the RYI employ
scaled-solar abundance ratios, whereas observations suggest that
Galactic globular clusters have an enhancement of $\alpha$-elements of
$[\alpha/Fe] \approx +0.4$ dex (e.g., \markcite{pt95}Pagel \&
Tautvai\v{s}ien\.{e} 1995).  \markcite{scs93}Salaris {\em et al.}
(1993) showed that for a given iron abundance, scaled-solar isochrones
0.29 dex more metal-rich in $[Fe/H]$ closely mimic $\alpha$-enhanced
isochrones with $[\alpha/Fe] = +0.4$ dex.  RYI with $[Fe/H] = -1.79$
indicate ages 1--2 Gyr older than those shown here.

Analogous RYI fits to the data of M68 (\markcite{wal93}Walker
1994) and M5 (\markcite{sbs96}Sandquist {\em et~al.} 1996), again
using the parameters from Table 2 and the ``$\alpha$-enhanced''
metallicities, produced ages of $\sim$13 Gyr for M68 and $\sim$12
Gyr for M5.  As in Sec. 3, M54 appears to be comparable in age to
these clusters, or perhaps slightly older.

We are concerned by the tendency for the isochrones to be bluer than
the data at $V \gtrsim 22$.  This could be due to (1) differential
incompleteness in our data as a function of color, (2) a tendency for
Sgr stars to dominate the red side of the lower main sequence and thus
to bias the data redward, or (3) inadequacies in the isochrones or
adopted reddening and distance modulus.  Though adopting a larger
reddening (e.g., by 0.05 mag $\approx 2\sigma$) corrects the main
sequence color problem and makes the derived age younger ($\sim$2
Gyr), it degrades the fit to the lower RGB.  Adjusting the reddening
and distance modulus in concert enables us to obtain a better overall
fit; the age obtained is 15 Gyr for $E(V-I) = 0.18$ and a distance
modulus 0.15 mag smaller than employed in Figure 3.

Our isochrone age estimates are supported by estimates based on the
luminosity of the subgiant branch.  The difference between the
magnitude of the subgiants at a well defined color and that of the
horizontal branch is similar for M54, M68, and M5.  When calibrated
using the RYI, we find that M54 is 1--2 Gyr older than M68 and 0--1
Gyr older than M5.  Details of this procedure are presented in Layden
\& Sarajedini (1997).

Given the uncertainties in determining absolute ages, we would like to
compare our age for M54 to that of the Sgr field population in a {\em
relative} sense.  In Figure 1, the lower RGB of the Sgr field
population appears to terminate abruptly at $V \approx 20.9$ mag.  RYI
with metallicities and ages with lower RGBs terminating at this
magnitude can be used to place an upper limit on the age of the Sgr
field.  For $[Fe/H] = -0.50$ (\markcite{sl95}SL95), we find a maximum
age of 6 Gyr.  For $[Fe/H] = -1.2$ and $[\alpha/Fe] = +0.4$, we find
an age of 9 Gyr.  For $[Fe/H] = -1.2$ and $[\alpha/Fe] = +0.0$, we
find an age of 11 Gyr.  The latter is the oldest age obtainable for
Sgr which is consistent with currently quoted abundance estimates.
This age is also in good agreement with the Sgr ages listed in Table
1.  Clearly, M54 must be older than the Sgr field stars by $\gtrsim$3
Gyr.

\section{Discussion}

All three of the methods discussed above suggest that M54 is 0.5--1.5
Gyr {\em older} that the comparison clusters M68 and M5.
\markcite{cds96}CDS find the age of M5 to be typical of Galactic
globular clusters of its metallicity, while M68 may be somewhat
younger than average.  Thus, M54 has an age typical of Galactic
globulars of its metallicity (see Figure 1 of \markcite{cds96}CDS).

The absolute age estimates discussed in Sec. 4, using the
\markcite{ldz90}LDZ relation between $M_V(RR)$ and $[Fe/H]$ and
$[\alpha/Fe] = +0.4$, suggest that M54 has an age of $\sim$14 Gyr.
Comparing this with the ages of the other Sgr globulars shown in Table
1 indicates that M54, Ter 8, and Arp 2 are all comparably old (for
more details, see \markcite{ls97}Layden \& Sarajedini 1997), while Ter
7 is significantly younger.  Given the uncertainties in the existing
photometry, we cannot rule out the possibility that the three old
clusters in Sgr are coeval.

Comparing our absolute age for M54 with the age estimates for the
dominant Sgr field population shown in Table 1 suggests that M54 is
older than the metal-rich field population in which it is embedded.
This result is supported by our analysis in Sec. 4, where we estimated
the maximum age of the Sgr field as a function of assumed $[Fe/H]$,
and found that M54 is at least 3 Gyr older than Sgr.

Taken at face value, these ages suggest that the metal-poor clusters
represent the earliest epoch of significant star formation in Sgr.
Vigorous star formation in the field appears to have begun several Gyr
later.  Given this age difference, it seems reasonable to expect that
gas expelled from evolving metal-poor cluster stars enriched the
interstellar medium and thus the first generation of Sgr field stars.
This explains, at least in part, why the Sgr field is so much more
metal-rich than the old clusters.  As was the case for many of the
Galactic satellite dwarf spheroidals (e.g.,
\markcite{tsh94}Smecker-Hane {\em et~al.}  1994), Sgr managed to
retain a significant portion of its gas for many Gyr, enabling the
formation of Ter 7, and of the $\sim$4 Gyr field population discussed by
\markcite{muskk}MUSKKK and represented by the blue plume of stars
above the MSTO in Figure 1.  Given the age and abundance of Ter 7, it
is perhaps more appropriate to compare this cluster with the
``populous clusters'' of the SMC or ESO121-SC03 in the LMC
(\markcite{gdc91}Da~Costa 1991) than with traditional globular
clusters.  Finally, we note that the HB morphologies of the three old
Sgr globulars are quite blue for their metallicity
(\markcite{sl95}SL95, \markcite{buo95}Buonanno {\em et~al.} 1995,
\markcite{og90}Ortolani \& Gratton 1990), in better agreement with the
Galactic globular clusters than those of the Fornax dwarf galaxy.  In
this respect, Sgr may be a better example of a surviving ``building
block'' of the Galactic halo than Fornax (see \markcite{rjz93}Zinn
1993).

\acknowledgments

We thank Mario Mateo for his thoughtful comments.  A.C.L. was supported
by NASA grant number HF-01082.01-96A, and A.S.  was supported by NASA
grant number HF-01077.01-94A, from the Space Telescope Science
Institute, which is operated by the Association of Universities for
Research in Astronomy, Inc. under NASA contract NAS5-26555.



\makeatletter
\def\jnl@aj{AJ}
\ifx\revtex@jnl\jnl@aj\let\tablebreak=\nl\fi
\makeatother

\begin{deluxetable}{cccc}
\tablewidth{22pc}
\tablecaption{The Ages of Sgr and its Clusters. \label{tbl-1}}
\tablehead{
\colhead{Object} & \colhead{$[Fe/H]$} &  \colhead{Age} &
\colhead{Source\tablenotemark{a}}
}
\startdata
Sgr  & --1.2 & 10   & 1  \nl
Sgr  & --0.5 & 12   & 2  \nl
Sgr  & --1.0 &10--14& 3  \nl
\medskip
Sgr  & --1.1 & 10   & 4  \nl
\medskip
Arp 2& --1.70&$12.3\pm0.8$ & 5  \nl
\medskip
Ter 7& --0.36& $7.2\pm0.5$ & 5  \nl
Ter 8& --1.99&$16.9\pm1.5$ & 5  \nl
\enddata
\tablenotetext{a}{Sources: (1)~MUSKKK, (2)~Mateo {\em et al.} (1996),
(3)~Fahlman {\em et al.} (1996), (4)~Marconi {\em et al.} (1997),
(5)~CDS.}
\end{deluxetable}


\makeatletter
\def\jnl@aj{AJ}
\ifx\revtex@jnl\jnl@aj\let\tablebreak=\nl\fi
\makeatother

\begin{deluxetable}{ccccc}
\tablewidth{30pc}
\tablecaption{Adopted Cluster Parameters. \label{tbl-2}}
\tablehead{
\colhead{Cluster} & \colhead{$V(HB)$} &
\colhead{$(V-I)_g$} &
\colhead{$[Fe/H]$} &  
\colhead{$E(V-I)$\tablenotemark{a}}
}
\startdata
M68 & 15.64 $\pm$ 0.01 & 0.979 &$-2.09 \pm 0.11$ & 0.09 \nl
M54 & 18.17 $\pm$ 0.05 & 1.095 &$-1.79 \pm 0.08$ & 0.17 \nl
M5  & 15.09 $\pm$ 0.02 & 0.966 &$-1.40 \pm 0.06$ & 0.04 
\enddata
\tablenotetext{a}{computed from $E(B-V)$ using the
relation of Dean {\em et~al.} (1978).}
\end{deluxetable}

\begin{figure}
\plotone{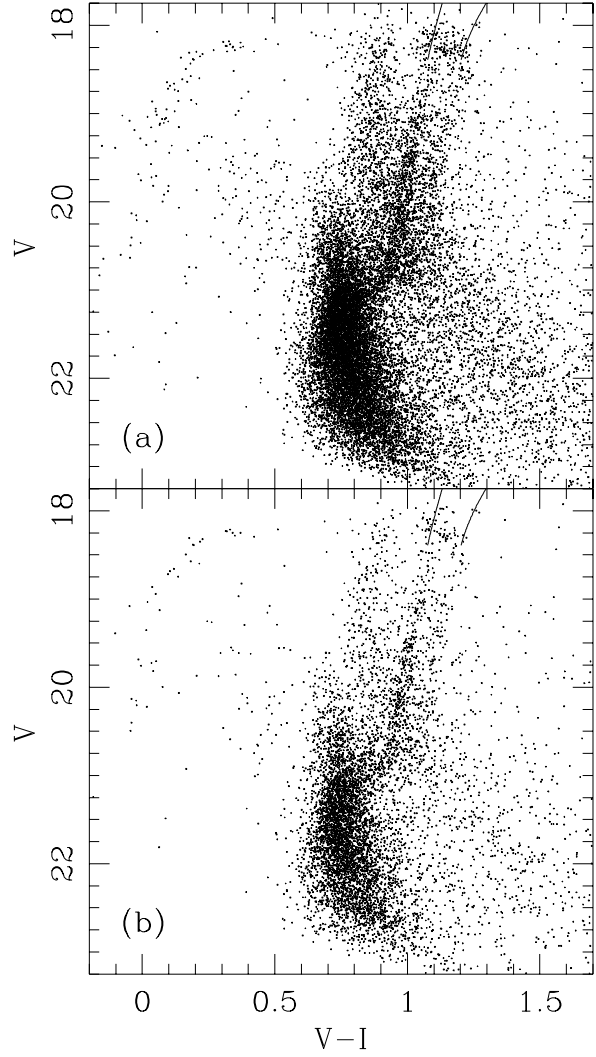}
\caption{(a) Color-magnitude diagram of 18,796 stars in the
direction of M54.  Only stars with at least three detections,
$\sigma_V < 0.050$ mag, and $\sigma_{V-I} < 0.071$ mag are shown.  (b)
As for (a), but only the 7551 stars located between 2.5 and 4.3 arcmin
from the cluster center are shown.  The curves are the red giant
branch fiducials of M54 (left) and Sgr (right) from SL95.
 \label{fig1}}
\end{figure}

\clearpage

\begin{figure}
\epsscale{1.01}
\plotone{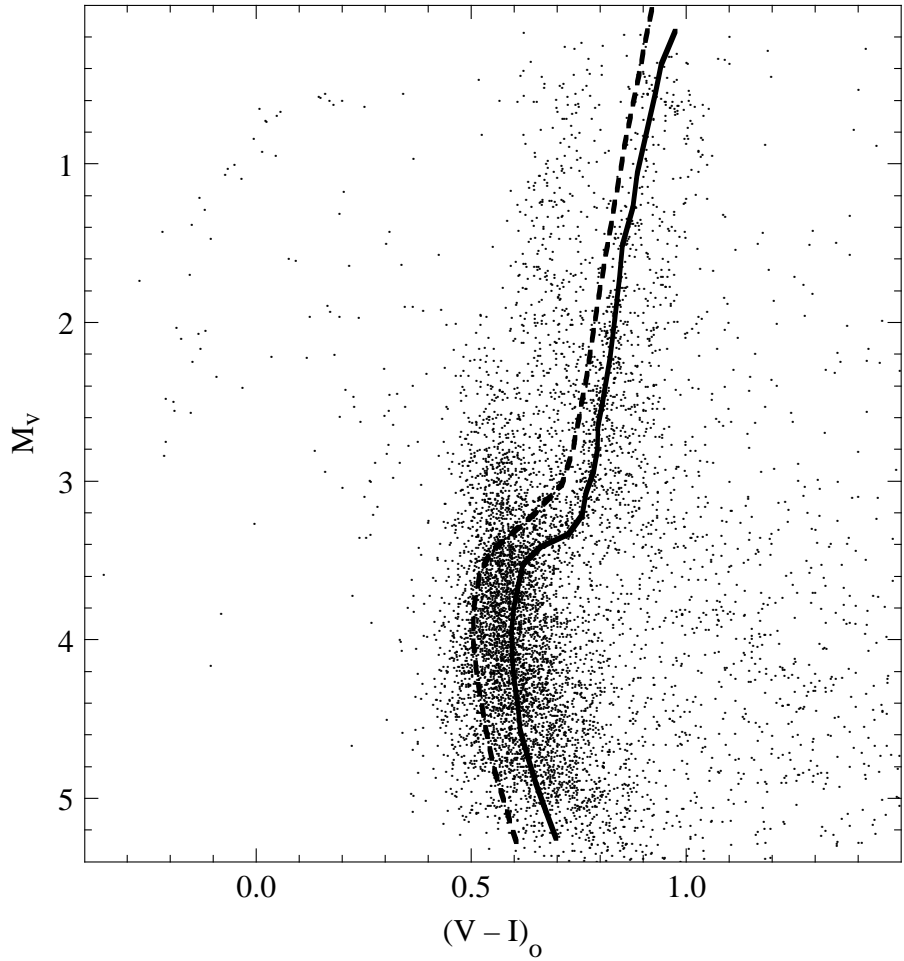}
\vskip -2.0in
\caption{As for Figure 1b, but the observed data have been shifted
to the theoretical plane as discussed in Sec. 3.  The solid line is
the fiducial for M5 ($[Fe/H] = -1.40$, Walker 1994) and the dashed
line is the fiducial for M68 ($[Fe/H] = -2.09$, Sandquist {\em et al.}
1996). \label{fig2}}
\end{figure}

\clearpage

\begin{figure}
\plotone{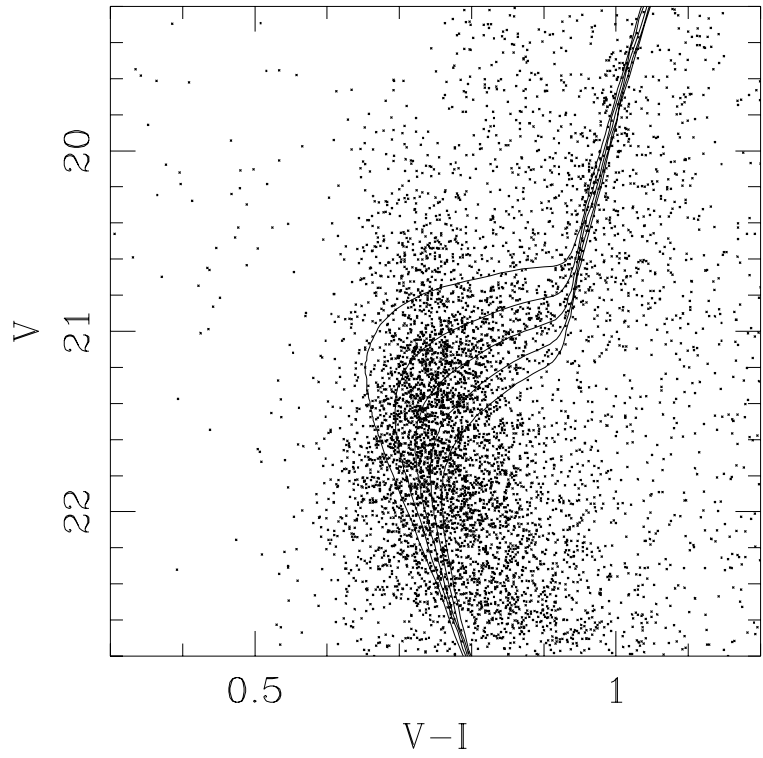}
\caption{The data from Figure 1b are plotted with Revised Yale
Isochrones shifted to the observed plane (see Sec. 4).  The isochrones
are for 10, 12, 14, 16, and 18 Gyr (left to right) and $[Fe/H] =
-1.50$.  The latter is equivalent to an isochrone with $[Fe/H] =
-1.79$ and $[\alpha/Fe] = +0.4$. \label{fig3}}
\end{figure}

\end{document}